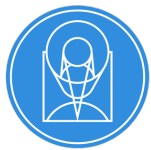

# JWST TECHNICAL REPORT

| Title: Characterization of the visit-to-visit Stability of the GR700XD Spectral Traces for NIRISS/SOSS Observations | Doc #: JWST-STScI-008448, SM-12<br>Date: 28 April 2023<br>Rev: - |
|---|---|
| Authors: Tyler Baines, Néstor Espinoza, Joseph Fillipazzo, and Kevin Volk  Phone: | Release Date: Release Date: 2 May 2023 |

## 1. Abstract

In this report, we present the results of our analysis of trace position changes during NIRISS/SOSS observations. We examine the visit-to-visit impact of the GR700XD pupil wheel (PW) position alignment on trace positions for spectral orders 1 and 2 using the data obtained to date. Our goal is to improve the wavelength solution by correlating the trace positions on the detector with the PW position angle. We find that there is a one-to-one correspondence between PW position and spectral trace rotation for both orders. This allowed us in turn to find an analytic model that is able to predict a trace position/shape as a function of PW position with sub-pixel accuracy of about ~0.1 pixels. Such a function can be used to predict the trace position in low signal-to-noise ratio cases, and/or as a template to track trace position changes as function of time in Time Series Observations (TSOs).

## 2. Introduction

The NIRISS/SOSS is one of the prime modes of JWST for exoplanet transit observations with its broad spectral range (0.8-2.8 µm) and moderate spectral resolution (R=700 at 1.4 µm) dispersed across three spectral orders. This observing mode has uncovered the very first set of exoplanet atmospheres with unprecedent detail as evidenced by recent studies (e.g., Pontoppidan et al., 2022; Feinstein, et al., 2023; Fu et al., 2022, Coulombe et al., 2023). However, while the instrument's performance within a single visit has been well characterized, both internally within the JWST project and by the scientific community, its visit-to-visit stability is still being carefully examined.

One of the visit-to-visit known issues is that the spectral traces (i.e., the path through a two-dimensional spectrum) of the GR700XD grism are observed to slightly rotate between visits. This can impact both automated spectral extraction algorithms and the wavelength solution of the extracted spectra. These rotations are thought to be caused by the difference between the commanded and actual PW position angles. When the pupil wheel rotates a wheel element into the optical path, the wheel does not cycle to the exact, commanded position. Instead, there is a slight offset. For the GR700XD, this offset can significantly impact the trace position on the detector for the SOSS observations by a few or more pixels. It's worth noting that while the GR700XD disperses light across three spectral orders, only orders 1 and 2 are supported for science operations at present, while order 3 will be supported in the future. Nonetheless, the focus of this report is on orders 1 and 2, which are sufficient for the purposes of investigating the visit-to-visit stability of the spectral traces. The work presented in this report sets the stage for developing a solution to improve the accuracy and stability of the wavelength solution between visits in future observations.

To illustrate one of the most striking symptoms of this mismatch, Figure 1 shows the order 1 traces and the spectrum of the star BD+60-1753 (the flux calibrator for NIRISS/SOSS used in commissioning) between Program ID (PID) numbers, 1091 and 1539. The deep absorption lines of this A-type dwarf star come from the Hydrogen Paschen series between 0.9-1.1 um. There is both a PW offset of ~0.07 degrees between the two trace samples (Figure 1, top) and a shift of ~1 pixel in the absorption lines (Figure 2, bottom), which is believed to arise from the trace rotation discussed above.

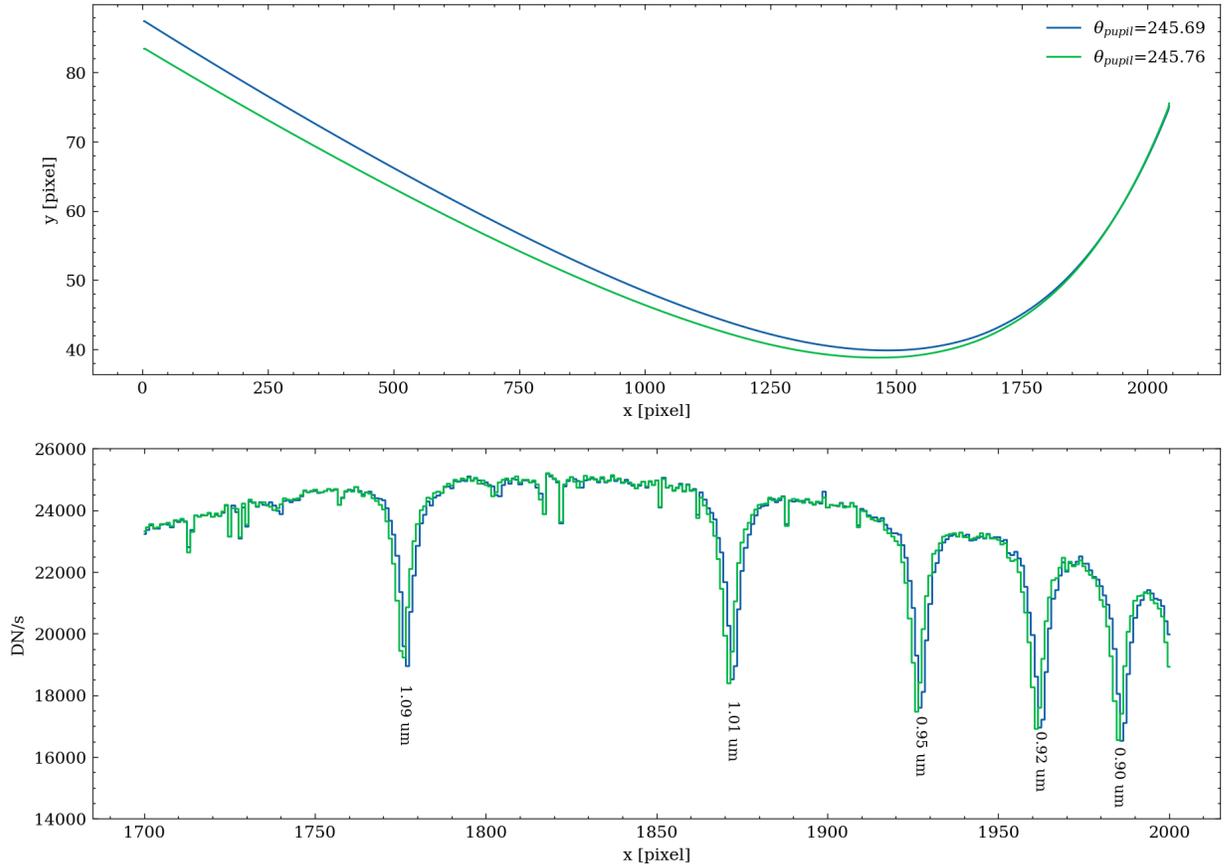

*Figure 1: An example demonstrating the influence the pupil wheel position has on the traces and extracted spectra of an A-type star, BD+60 1753, observed in SOSS mode from a commissioning program (PID 1091, green) and calibration program (PID 1539, blue). Between the observations, an offset of about 0.07 degrees is seen between the PW positions, as well as a shift in the hydrogen lines by ~1 pixel.*

By characterizing how the geometric properties of the spectral traces change at different PW positions this Technical Report aims to enhance our ability to predict the trace's behavior for any given PW position in future observations. This analysis will serve as the basis for future predictive models that can identify changes in the wavelength solution of offset traces, which will be investigated in a future Technical Report. In cases where obtaining a precise measurement of the traces becomes challenging, e.g., very faint sources, the predictive model will enable users to still acquire accurate spectra from such sources with NIRISS/SOSS.

### 3. Data Collection and Processing

We compiled a comprehensive list of all available SOSS observing programs from 2022 June to 2022 December, including those used in commissioning, calibration, and science programs, to perform our data analysis. Given the limited number of SOSS observations, we gathered as many datasets as possible by leveraging both public and proprietary data from the MAST (Mikulski Archive for Space Telescopes); see Table 1. We acquired level 2a data products, which are the rates per integration (i.e., the _rateints.fits files) resulting from the STScI/JWST pipeline reduction (Bushouse et al. 2022). All observations used the CLEAR+GR700XD configuration with the SUBSTRIP256 subarray.

**Table 1: List of the SOSS observation programs along with the program PIs, number of observations in each program with their respective measured pupil wheel positon angles.**

| Program ID | Program PI | Number of Observations | $\theta_{pwcpos}$ [deg] |
|---|---|---|---|
| 1091 | André Martel | 1 | 245.791 |
| 1201 | David Lafrenière | 3 | 246.666, 245.693, 245.740 |
| 1366 | Natalie Batalha | 1 | 245.718 |
| 1536 | Karl Gordon | 4 | 245.825, 245.906, 245.918, 245.293 |
| 1538 | Karl Gordon | 1 | 245.801 |
| 1539 | Karl Gordon | 3 | 245.688, 245.757, 245.879 |
| 1541 | Néstor Espinoza | 1 | 245.898 |
| 2589 | Olivia Lim | 3 | 245.920, 245.859, 245.681 |
| 2734 | Klaus Pontoppidan | 2 | 245.886, 245.798 |

We measure the median count rate image of all integrations in a given exposure to obtain a high signal-to-noise ratio image for our analysis. We also extract the actual PW position angle stored in the FITS header using the "PWCPOS" keyword. Figure 2 shows the PW angles as a function of observation date. We measure the mean and standard deviation of the position angles and find a mean value of $\boldsymbol{\theta_{pwcpos} = 245.8081 \pm 0.0911}°$, which is different than the commanded position of $\boldsymbol{\theta_{cmd} = 245.7600}°$ (Martel et al. 2021). All reported PW position angles fall within an acceptable tolerance range of $\pm 0.1651°$ from the commanded position. Furthermore, we also observe a bimodal distribution of the position angles, aligning with the findings of Martel et al. (2016, 2021, 2022) from ground testing and the most recent on-orbit wheel behavior. This bimodality originates from the PW's direction of approach, clockwise or counter-clockwise, which leads to negative and positive offsets respectively.

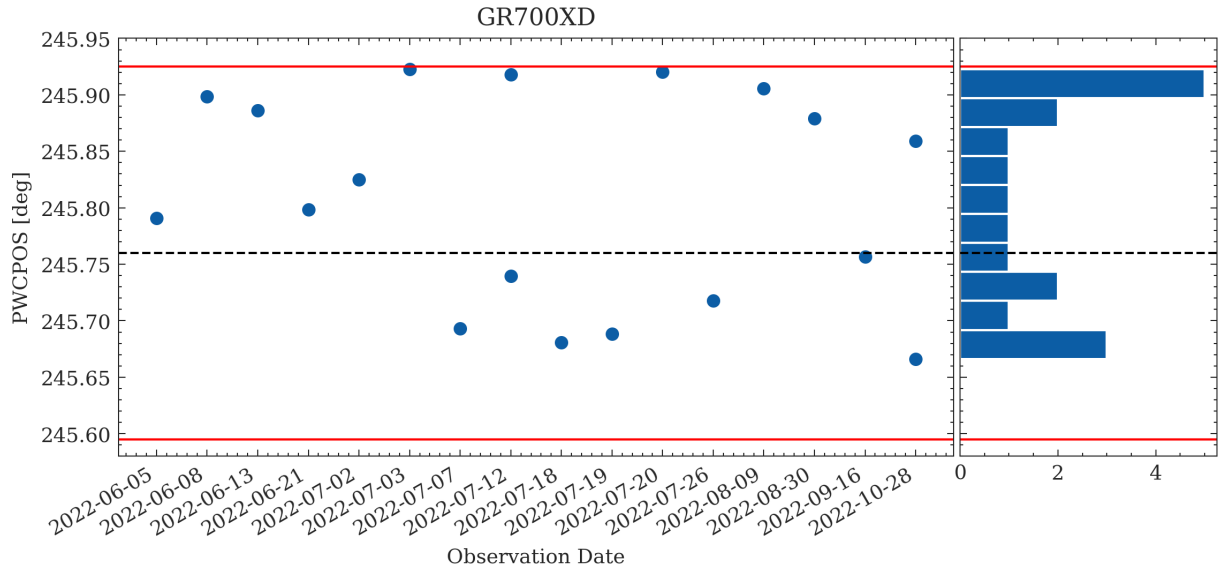

*Figure 2: The GR700XD pupil position angles extracted from the "PWCPOS" keyword as a function of observation date. The solid dashed line represents the command position angle and the red lines the uncertainties on this position $\pm 0.1651°$, see Martel et al. (2016, 2021).*

## 4. Spectral Tracing and Data Analysis

We used the *transitspectroscopy* package (Espinoza, 2022) to compute the traces for each of our median frames. The code performs a cross-correlation on each column of the detector using a double gaussian kernel with a 15-pixel separation and 3-pixel standard deviation to estimate the trace centroids. These are then smoothed using a B-spline with knots defined on empirically determined portions of the trace for each order (see Feinstein et al., 2023 for more details). To ensure good quality samples, we visually inspected the traces and rejected any that showed poor performance, likely due to contamination from other dispersed sources in the field of view. Most of the order 1 traces were of good quality, but we reject a few samples from order 2. To refine the order 2 traces, we ignored detector pixels with column values below 1000 due to some variability. Program 1538's measured traces were rejected entirely because of strong source contamination across each spectral order.

For each order, we compare the traces as a function of PW position angle. We identified the pivot points at which the traces appear to rotate about, denoted by the red dot as shown in Figure 3. The traces at bluer wavelengths (larger x-pixel positions) are closer together and deviate at the red end (smaller x-pixel positions). The exact cause of this behavior is unclear; however, we hypothesize that this might be due to distortion corrections which are known to be wavelength- and position-dependent. The pivot points were found by measuring the scatter of the traces y-positions along each column and identifying a point at which the scatter was at a minimum. We find the integer pixel coordinates of the pivot positions to be located at (x=1887, y=54) and (x=1667, y=200) for orders 1 and 2 respectively. This approach provides a rough approximation of the trace's center of rotation.

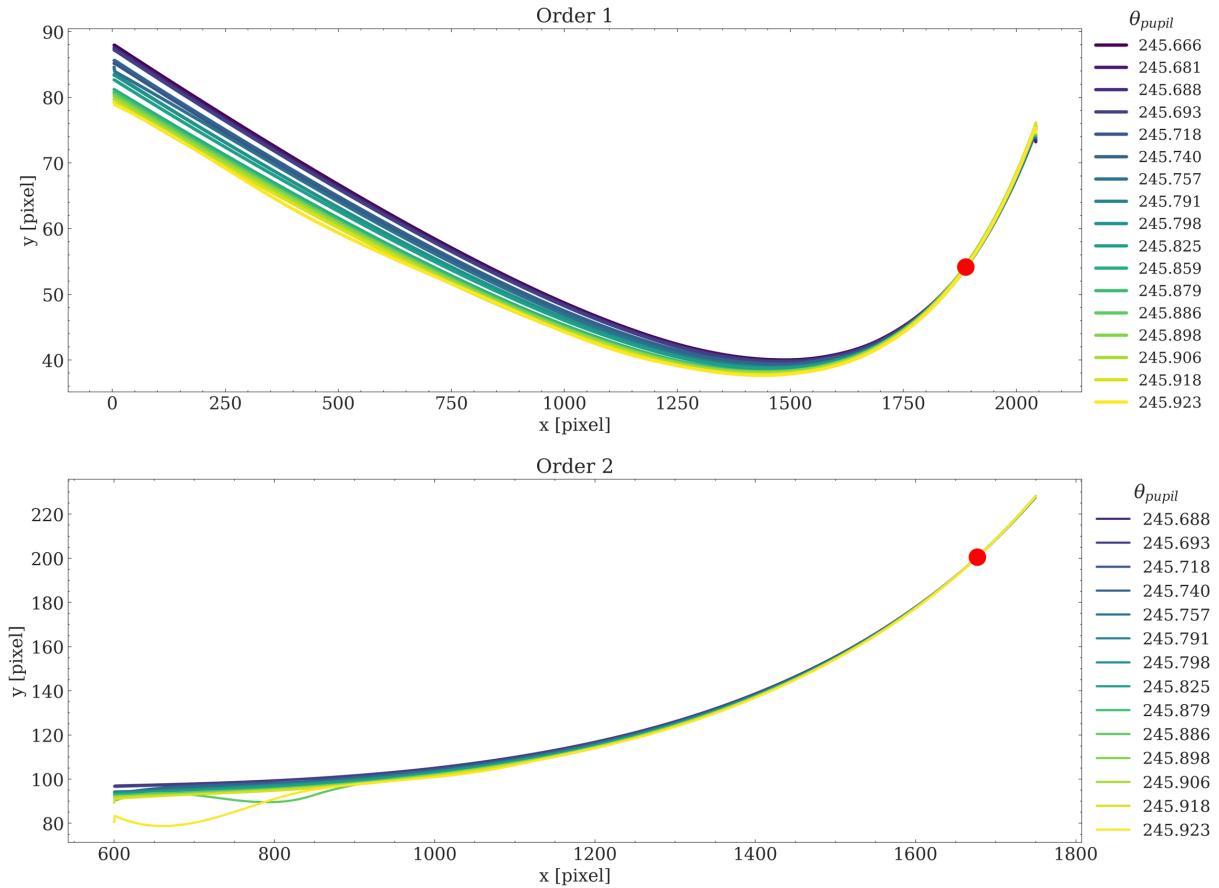

*Figure 3: Measured trace positions for order 1 and 2. Order 1 traces are well-behaved while the order 2 trace positions between detector columns 600 and 1000 had difficulties due to source contaminations. The color gradient shows how the traces change with the pupil wheel position angle. The red dot is the pivot point of the traces.*

We then we measured the trace rotation angle for each of our trace samples and compared it against the PW position angle. The rotation angle $\theta_{rot}$ is the angle between the two lines that connect the edges of the trace to its pivot point and is found by using basic geometry (as illustrated in Figure 4) where $\theta_{rot} = 180° - (\theta_L + \theta_R)$, $\theta_L$ and $\theta_R$ are the angles of the left and right edges with respect to the trace's center of rotation. It appears the relationship between the rotation angle and PW position angle is linear (see Figure 5), which suggests that the trace rotation follows the PW position closely, and thus the latter might be an excellent predictor of the trace shape. We find that our result aligns with what was found in Martél et al (2016; see their Figure 7).

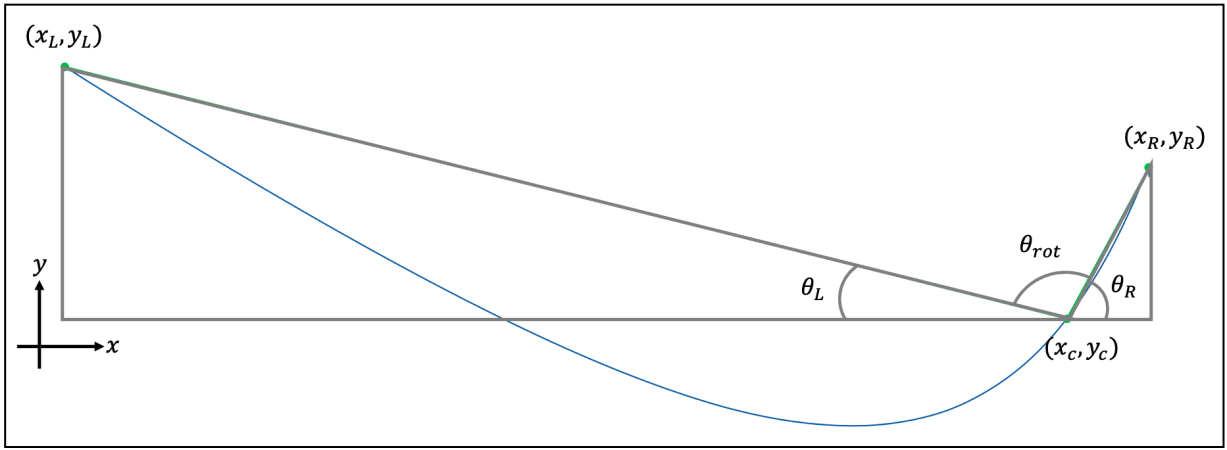

*Figure 4: Illustration of the geometry to determine the rotation angle of the trace (blue) between the edges with respect to the pivot point. The angles $\theta_L$, $\theta_R$, and $\theta_{rot}$ are supplementary to a straight line parallel to the x-axis that intersects the trace at $(x_c, y_c)$. By knowing the values at the trace's edges and the pivot point, we can determine the angles $\theta_L$ and $\theta_R$, and subsequently, find $\theta_{rot}$.*

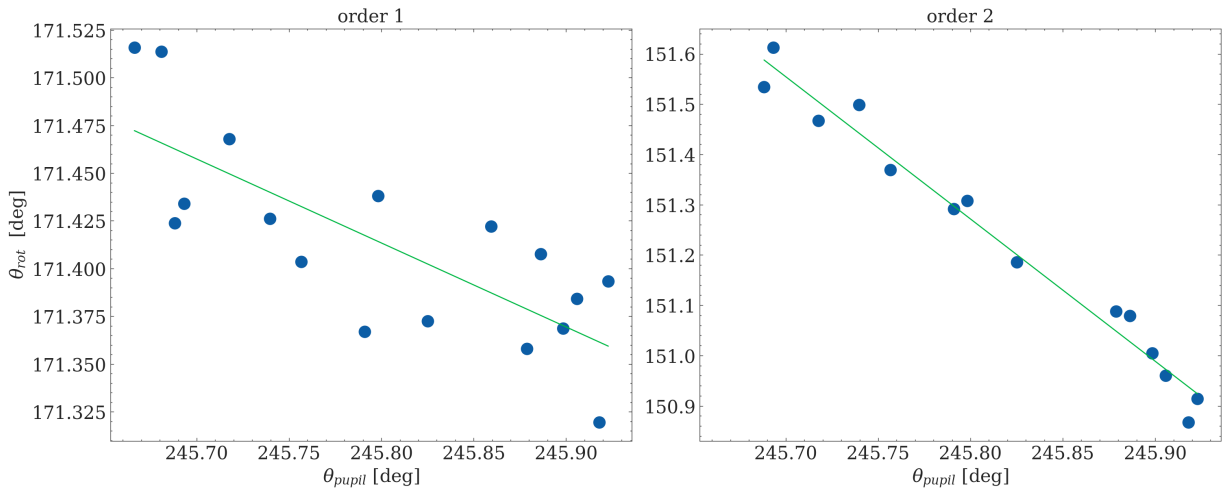

*Figure 5: Calculated rotation angles as a function of pupil position angle for orders 1 and 2.*

Next, we apply a linear regression along each x-pixel column of the trace samples by fitting the y-positions as a function of position angle with respect to the commanded position (i.e., $y(\theta, x) = m(x)(\theta - \theta_{cmd}) + b(x)$ ). By doing so, it is possible to predict the trace y-position at a given position angle along the spatial x-direction. Each column will have its own set of coefficients $(m, b)$ such that the slope $m$ and intercept $b$ are functions of $x$. The function $b(x)$ describes the trace y-positions of the commanded position and can serve as the reference trace. We repeat this process for the order 2 traces.

Now that we have a reference model, we attempted to de-rotate the trace samples and align them to the model. A user should not de-rotate the data in order to match the traces from a calibrated data product. Instead, one should de-rotate a reference model to predict the trace positions to the associated PWCPOS in a given observation. Thus, we apply a rotation transformation to the $i^{th}$ measured trace sample position $(x_i, y_i)$ using the respective pivot point for a given spectral order as the center or rotation $(x_c, y_c)$. The angle $\alpha_i$ used to rotate the trace is determined by taking the difference between the corresponding measured pupil position angle $\theta_i$ and the commanded position (i.e., the offset, $\alpha_i = \theta_i - \theta_{CMD}$). Thus, the rotation transformation is given by,

$$\begin{bmatrix} x_{i,new} \\ y_{i,new} \end{bmatrix} = \begin{bmatrix} \cos \alpha_i & -\sin \alpha_i \\ \sin \alpha_i & \cos \alpha_i \end{bmatrix} \begin{bmatrix} x_i - x_c \\ y_i - y_c \end{bmatrix} + \begin{bmatrix} x_c \\ y_c \end{bmatrix}.$$

We apply our coordinate transform method to each of our traces and interpolate the new positions onto the original x-coordinates. The resulting de-rotated traces and residuals for order 1 are shown in Figure 6. The de-rotated traces show residuals of less than 1 pixel at most, with the majority being less than 0.5 pixels, compared to the reference trace positions. We observed biases in our residuals which may suggest that the center of the rotation that we have chosen might not be the optimal position. It might be possible that the center of rotation is not on the trace at all. In addition, our residuals show oscillatory characteristics which is due to the contamination of overlapping spectral orders from nearby sources causing the measured trace to deviate from its expected position.

For future observations, this reference trace of the commanded position and rotation transform will be used to predict the trace position at any given pupil position angle at the sub-pixel level and be robust against signal contamination. The supporting code and models for predicting the trace positions for a given PW position angle is released in the package PASTASOSS

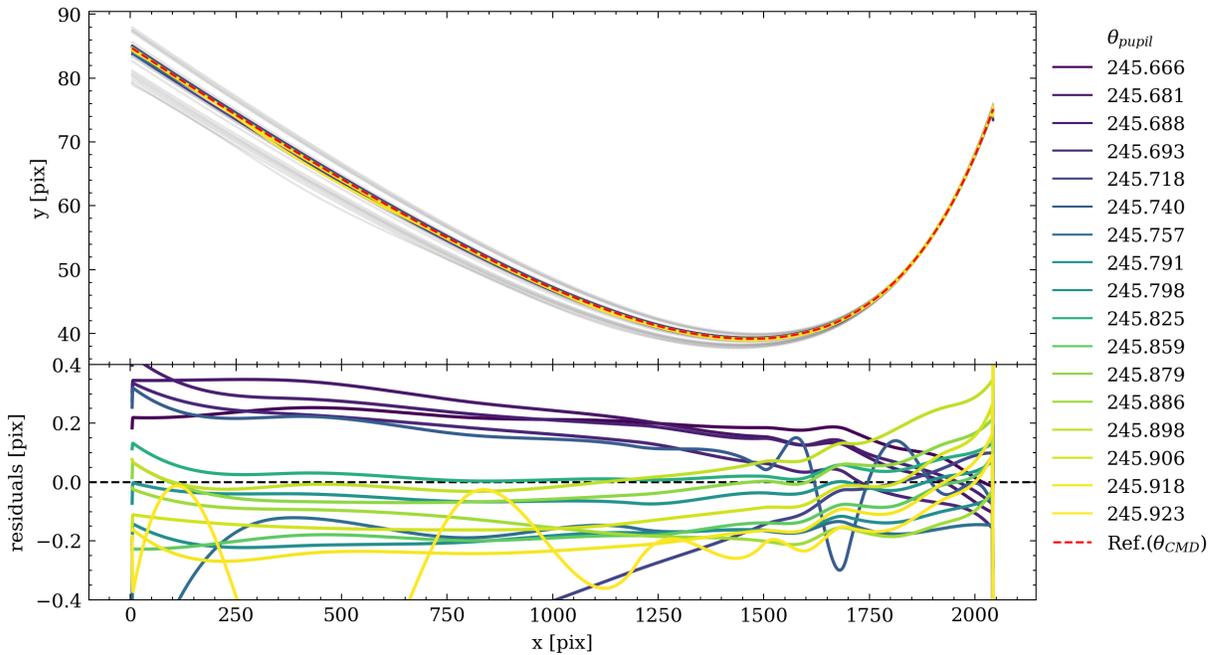

*Figure 6: The top panel displays the order 1 spectral traces before (grey lines) and after (colored lines) de-rotation to the reference position angle of $\theta_{cmd} = 245.7600°$. The bottom panel shows the corresponding residuals, where features with residuals greater than 0.5 pixel are attributed to source contamination while the residual slope may indicate an offset in the center of rotation.*

(**P**redicting **A**ccurate **S**pectral **T**races for **A**strophysical **SOSS** Spectra) can be found on GitHub [https://github.com/spacetelescope/pastasoss].

## 5. Conclusion

Using all the currently available NIRISS SOSS observations from 2022 June to 2022 December, we have measured the traces for spectral orders 1 and 2. For each order, we find that the traces appear to rotate about their own respective pivot points in a counter-clockwise direction. In addition, the rotational angle of the traces is found to be linearly correlated with the PW position angle (i.e., the PW position predicts the rotation angle) which is in agreement from results found during ground and in-flight testing. We have shown that de-rotating the traces about their pivot point yields sub-pixel performance. We now have a model (PASTASOSS) that predicts the trace positions for any given PW position angle. PASTASOSS can be run as a standalone analysis tool, with implementation into the JWST pipeline a possibility for future enhancements. Having a well-aligned trace is important as it provides the foundation to solve the wavelength calibration as a function of PW position angle.

## 6. Acknowledgements

We thank Jo Taylor and André Martel for reviewing and providing useful comments for this report.

## 7. References

Coulombe, L-P., Benneke, B., Challener, R. et al, 2023, "A broadband thermal emission spectrum of the ultra-hot Jupiter WASP-18b", eprint arXiv:2301.08192

Espinoza, N., 2022. "TransitSpectroscopy" (0.3.11). Zenodo. https://doi.org/10.5281/zenodo.6960924

Feinstein, Adina D., et al. "Early Release Science of the exoplanet WASP-39b with JWST NIRISS." Nature (2023)

Fu, G., Espinoza, N., Sing, D. et al., 2022, "Water and an Escaping Helium Tail Detected in the Hazy and Methane-depleted Atmosphere of HAT-P-18b from JWST NIRISS/SOSS", The Astrophysical Journal Letters, Volume 940, Issue 2, id.L35, 8 pp.

Pontoppidan, K., Barrientes, J., Blome, C. et al., 2022, "The JWST Early Release Observations", The Astrophysical Journal Letters, Volume 936, Issue 1, id.L14, 13 pp.

Martel, André R. "Analysis of NIRISS CV3 Data: The Repeatability of the Pupil Wheel and the Filter Wheel." Technical Report JWST-STScI-004964 (2016): 4964.

Martel, André R. "Analysis of NIRISS OTIS Data: The Repeatability of the Pupil Wheel and the Filter Wheel." Technical Report JWST-STScI-006035 (2021): 6035.

Martel André R. "The Early Behavior if the NIRISS Pupil Wheel and Filter Wheel", Technical Report JWST-STScI-008298 (2022): 8298.

Bushouse, Howard, Eisenhamer, Jonathan, Dencheva, Nadia, Davies, James, Greenfield, Perry, Morrison, Jane, Hodge, Phil, Simon, Bernie, Grumm, David, Droettboom, Michael, Slavich, Edward, Sosey, Megan, Pauly, Tyler, Miller, Todd, Jedrzejewski, Robert, Hack, Warren, Davis, David, Crawford, Steven, Law, David, … Jamieson, William. (2022). JWST Calibration Pipeline (1.7.0). Zenodo. https://doi.org/10.5281/zenodo.7038885